\documentclass[superscriptaddress,twocolumn,floats,showpacs,prl,amsmath,amssymb,floatfix,nofootinbib,balancelastpage]{revtex4}

\usepackage{graphicx}

\newcommand{\bea}{\begin{eqnarray}}
\newcommand{\eea}{\end{eqnarray}}
\newcommand{\bean}{\begin{eqnarray*}}
\newcommand{\eean}{\end{eqnarray*}}

\newcommand{\bk}{{\bf k}}

\begin{document}

\title{Impact of Secondary non-Gaussianities on the Search for Primordial Non-Gaussianity with CMB Maps}
\author{Paolo Serra and Asantha Cooray}
\affiliation{Department of Physics and Astronomy, 4186 Frederick Reines Hall, University of California, Irvine, CA 92697}

\begin{abstract}
When  constraining the primordial non-Gaussianity parameter $f_{\rm NL}$ with
cosmic microwave background anisotropy maps, the bias resulting from the covariance between
primordial non-Gaussianity and secondary non-Gaussianities to the estimator of $f_{\rm NL}$ is generally assumed to be negligible.
We show that this assumption may not hold when attempting to measure the primordial non-Gaussianity 
out to angular scales below a few tens arcminutes with an experiment like Planck, 
especially if the primordial non-Gaussianity parameter  is around the minimum detectability level
with $f_{\rm NL}$ between 5 and 10. In future, it will be necessary to jointly estimate the combined primordial and secondary
contributions to the CMB bispectrum and establish $f_{\rm NL}$ by properly accounting for the 
confusion from secondary non-Gaussianities.
\end{abstract}


\maketitle

\noindent \emph{Introduction--- } 
The search for primordial non-Gaussianity with constraints on the non-Gaussianity parameter $f_{\rm NL}$  
using cosmic microwave background (CMB) anisotropy maps is now an active topic in
cosmology today \cite{Komatsu,Lig,Komatsu4}. The 3-year Wilkinson Microwave Anisotropy Probe has allowed the constraint that
$-54 < f_{\rm NL} <114$ at the 95\% confidence level \cite{Spergel}, though a more recent study claims a non-zero detection  of primordial non-Gaussianity
at the same 95\% confidence level with $26.9 < f_{\rm NL} < 146.7$ \cite{Yadav}.  This 
result, if correct,  has significant
cosmological implications since the expected value under standard inflationary models is  $f_{\rm NL} \lesssim 1$  \cite{Salopek,Falk,Gangui,Pyne,Acquaviva,Maldacena,Bartolo}, though alternative models of inflation, such as the ekpyrotic cosmology \cite{Buchbinder,Lehners}, 
generally predict a large primordial non-Gaussianity with $f_{\rm NL}$  at few tens.

Most studies that constrain $f_{\rm NL}$ with CMB anisotropy maps make use of an estimator for $f_{\rm NL}$ 
of the form \cite{Komatsu2,Creminelli,Yadav2}
\begin{equation}
\hat{f}_{\rm NL} = \frac{\hat{S}_{\rm prim} + \hat{S}^{\rm lin}}{N}\, ,
\end{equation}
where
\begin{equation}
\hat{S}_{\rm prim} = \sum_{l_1 l_2 l_3} \frac{B^{\rm prim}_{l_1l_2l_3} \hat{B}^{\rm obs}_{l_1 l_2 l_3}}{\sigma^2(l_1,l_2,l_3)} \, ,
\label{eqn:S}
\end{equation}
when $B^{\rm prim}_{l_1 l_2 k_3}$ is the primordial bispectrum with the assumption that $f_{\rm NL}=1$.
Here, $\hat{S}^{\rm lin}$ is a linear correction to account for issues related to the maps (such as the mask) 
and $N$ is an overall normalization factor \cite{Yadav2}.
For the present discussion motivated from an analytical calculation, we can ignore the correction associated with $\hat{S}^{\rm lin}$ 
which involves imperfections in the data, such as due to the mask. We also assume all-sky data here.
In equation~(\ref{eqn:S}), $\sigma^2$ is the noise variance to the bispectrum \cite{Komatsu,Cooray}.

In general $\hat{B}^{\rm obs}_{l_1 l_2 l_3} = f_{\rm NL} B_{l_1 l_2 l_3}^{\rm prim} + b_{\rm ps} B_{l_1 l_2 l_3}^{\rm ps} + A_{\rm SZ} B_{l_1 l_2 l_3}^{\rm SZ-\kappa} +A_{\rm ISW} B_{l_1 l_2 l_3}^{\rm ISW-\kappa}+...$,
where $B_{l_1 l_2 l_3}^{\rm ps}$ is the shape of the non-Gaussianity with an overall normalization 
given by $b_{\rm ps}$ and $B_{l_1 l_2 l_3}^{\rm SZ-\kappa}$ and  $B_{l_1 l_2 l_3}^{\rm ISW-\kappa}$ are additional
foreground, secondary non-Gaussianities from the SZ and ISW effects correlating with CMB lensing \cite{Cooray,Goldberg}.
These are certainly not all the non-Gaussian contributions to a CMB map.
There are non-Gaussianities from
ISW \cite{Cooray3}, kinetic SZ/Ostriker-Vishniac \cite{Cooray4}, and the SZ effect itself \cite{Cooray2}. 
We ignore ISW and kinetic SZ/OV related bispectra as they
are small compared to SZ generated non-Gaussianities.  
The SZ-SZ-SZ bispectrum is significant
at arcminute angular scales, but given the power-law shot-noise behavior of the SZ bispectrum when $l < 1500$, 
the SZ contribution to the bispectrum can be thought of as
an additional correction to $b_{\rm ps}$. The shot-noise behavior of the SZ effect is especially applicable for
the SZ contribution during reionization associated with hot electrons in supernovae bubbles Compton-cooling 
off of the CMB \cite{Oh}. Thus, we do not separately include the total SZ bispectrum as a separate non-Gaussianity here.

When estimating $f_{\rm NL}$, it is usually assumed that
$\hat{B}^{\rm obs}_{l_1 l_2 l_3} \approx f_{\rm NL} B_{l_1 l_2 l_3}^{\rm prim}$ when estimating the primordial 
bispectrum.  This allows an estimator for  $f_{\rm NL}$  through
\begin{equation}
\hat{S}_{\rm prim} = f_{\rm NL} \sum_{l_1 l_2 l_3} \frac{\left(B^{\rm prim}_{l_1l_2l_3}\right)^2}{\sigma^2(l_1,l_2,l_3)} \, ,
\label{eqn:fnl}
\end{equation}
with $\hat{f}_{\rm NL}=\hat{S}_{\rm prim}/N$, where the normalization $N$ is simply the summed term.
The above assumption  that only the primordial non-Gaussianity can be considered is generally motivated 
by the fact that the covariance term associated with the mode 
overlap between $B_{l_1l_2l_3}^{\rm prim}$ and additional secondary contributions to $B_{l_1l_2l_3}^{\rm obs}$
via
\begin{equation}
\hat{S}_{prim,cov} = \sum_i A_i \sum_{l_1 l_2 l_3} \frac{B^{\rm prim}_{l_1l_2l_3}B^{i}_{l_1l_2l_3}}{\sigma^2(l_1,l_2,l_3)} \, ,
\label{eqn:sec}
\end{equation}
when $A_i=(b_{\rm ps}, A_{\rm SZ}, A_{\rm ISW},...)$ 
is expected to be smaller than the dominant term 
from equation~(\ref{eqn:fnl}) \cite{Komatsu}. Nevertheless, an estimate of $f_{\rm NL}$ only from equation~(\ref{eqn:fnl}) leads to a biased estimate
because of the contributions from secondary  anisotropies through equation~(\ref{eqn:sec}).

\begin{figure}[t]
\includegraphics[scale=0.38,angle=-90]{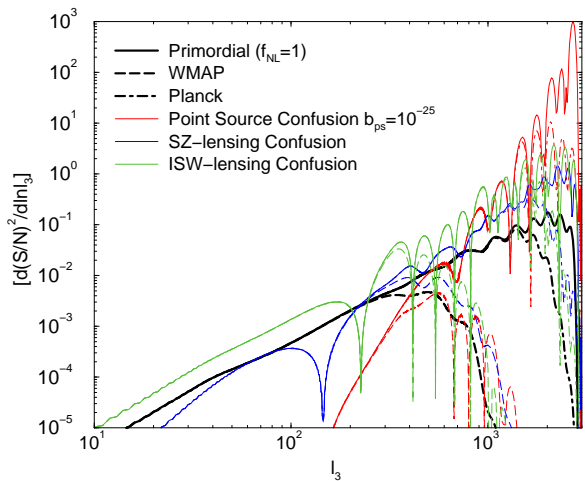}
\caption{Absolute values of signal-to-noise ratio squared for the detection of primordial bispectrum (black lines) assuming $f_{\rm NL}=1$ as a function of $l_3$. 
The signal-to-noise ratio squared for WMAP and Planck are shown with dashed and dot-dashed line, respectively. The red, blue, and green lines show the confusion resulting from
the covariance between primary and point source, primary and SZ-lensing, and primary and ISW-lensing 
bispectra, respectively.
}
\label{cl}
\end{figure}

While the CMB map contains a large number of secondary non-Gaussian signals,
in terms of the covariance related to the $f_{\rm NL}$ measurement, 
what is necessary is not to account for all of these non-Gaussianities, but to account for non-Gaussianities with
bispectrum shapes $B_{l_1l_2l_3}$ in $(l_1,l_2,l_3)$ moment space 
that align with the shape of the primary bispectrum. In this respect,
previous calculations have suggested that the point-source bispectrum may be
ignored \cite{Komatsu}, but the ISW-lensing bispectrum must be accounted for the Planck analysis \cite{Smith}. 

Including the SZ-lensing bispectrum, we find that while the assumption that the covariance from 
secondary anisotropies can be mostly ignored for an experiment like WMAP,
it may be necessary to account for certain covariances when 
estimating $f_{\rm NL}$ from a high resolution experiment like Planck, especially if the underlying primordial
non-Gaussianity has a value around $f_{\rm NL}$ between 5 and 10 consistent with
the minimum amplitude detectable with Planck. At the minimum detectability level of
WMAP with $f_{\rm NL} \sim 20$, the secondary 
anisotropies involving both residual points sources and lensing correlations
will bias $f_{\rm NL}$ by a factor between 1.2 and 1.5 if
primordial non-Gaussianity estimate is performed out to 
angular scales corresponding to $\ell > 700$.

To reach these conclusions, we first calculated 
$B^{\rm prim}_{l_1l_2l_3}$ following Ref.~\cite{Komatsu} with the full radiation transfer function using
a modified code of CMBFAST \cite{Seljak} for the standard flat $\Lambda$CDM cosmological model 
consistent with WMAP with $\Omega_b=0.042$, $\Omega_c=0.238$, $h=0.732$, $n=0.958$, and $\tau=0.089$. 
We verified our calculations are consistent with prior calculations in the literature. 
In Fig.~1, we show the
the absolute value of the signal-to-noise square ratio for the primordial bispectrum (thick lines) and 
for the covariances between primary and secondary bispectra. 
The plotted quantity here involving $d(S/N)^2/d\ln l_3$ resembles
the estimator $\hat{S}$ above, expect for the sum over $l_3$ while keeping the sign (ignoring the sign changes lead to a higher bias
as described in Ref.~\cite{Smith}).
While the primordial calculation involves $\sum (B^{\rm prim})^2/\sigma^2$, the
``signal-to-noise'' square of the covariance follows from $b_{\rm ps} \sum B^{\rm prim} B^{\rm ps}/\sigma^2$, for example for
the point-source confusion, and these confusions should not be interpreted simply as the
signal-to-noise ratio square to detect any of these secondary bispectra directly from the CMB maps. 

Instead of squared signal-to-noise ratios, 
to highlight the bias introduced to $f_{\rm NL}$ when the estimator ignores secondary non-Gaussianity covariances, we calculated
$f_{\rm NL}^{\rm tot} = f_{\rm NL} + f_{\rm bias}$ where $f_{\rm bias}$ is the bias that is generated artificially by the  correlation
of modes between the primordial bispectrum and secondary bispectra. 
To properly normalize the relative contribution from secondary non-Gaussianities,
we assume normalizations for the point-source bispectrum consistent with WMAP with $b_{\rm ps}=3 \times 10^{-25}$,
consistent with Q+V+W residual foreground \cite{Spergel}, and Planck with
$b_{\rm ps}=5 \times 10^{-27}$. The $b_{\rm ps}$ value for Planck is slightly higher than the values routinely quoted in the literature for
unresolved radio sources in Planck high resolution maps, but this is due to the fact that we believe $b_{\rm ps}$ includes additional contributions
such as from the SZ-SZ-SZ bispectrum from both clusters at low redshifts and supernovae halos during reionization 
with a power-law shot-noise spectrum when $l < 1500$.
For the ISW-lensing and SZ-lensing bispectrum, we follow the calculation of Ref.~\cite{Cooray} and  generate the
SZ contribution and the SZ correlation with dark matter halos responsible for lensing of the CMB using the halo model \cite{Cooray2}.
To account for an overall uncertainty and the variation in SZ and ISW amplitudes
we have introduced an overall amplitude $A_{\rm SZ}$ and $A_{\rm ISW}$ respectively. 
Finally, to illustrate our results, we assume $f_{\rm NL}$ consistent with roughly the minimum detectable primordial non-Gaussianity with WMAP and
 Planck with $f_{\rm NL}=20$ and $5$, respectively. As we find later, the dominant confusion is from lensing bispectra and not
from point sources.

\begin{figure}[t]
\includegraphics[scale=0.38,angle=-90]{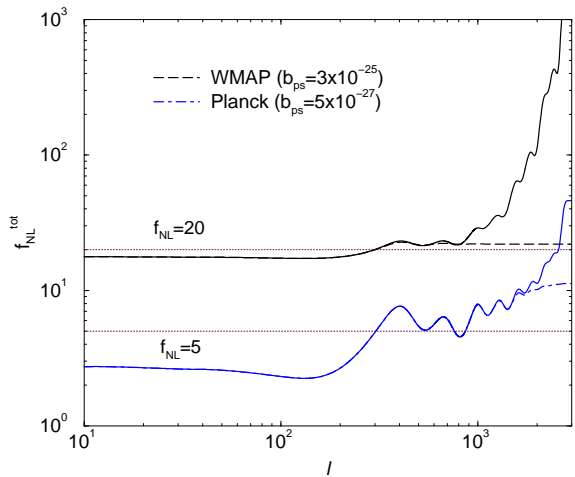}
\caption{The maximum non-Gaussianity measured with an optimal estimator for the primordial bispectrum $f_{\rm NL}^{\rm tot}$, which includes the true underlying
primordial non-Gaussianity with $f_{\rm NL}$ as labeled on the figure and the bias correction coming from the unaccounted secondary anisotropies.
The bias is generally small and  non existing if primordial non-Gaussianity measurements are limited out to $l < 500$, but depending on the value of $f_{\rm NL}$ and the residual
point source contamination, the correction is generally a factor of 1.5 to 2. If $f_{\rm NL} \lesssim 10$, for Planck,  it is necessary to account for secondary non-Gaussianities properly.
}
\label{fnl}
\end{figure}

We summarize our results in Fig.~2, where we plot $f_{\rm NL}^{\rm tot}$ which
can be thought of as the total primordial non-Gaussianity parameter that one will extract with the above estimator for $f_{\rm NL}$ when
no attempt has been made to separate out the confusion  from secondary anisotropies. For the most part, the bias is negligible
and becomes only important when $l > 500$. For WMAP, shown with a dashed line in Fig.~2 with the assumption that $f_{\rm NL}=20$
if non-Gaussianity measurements are attempted out to $l > 700$, capturing basically all information in WMAP maps,
then one finds a bias between a factor of 1.5 to 2 if $f_{\rm NL} \sim 20$. If $f_{\rm NL} > 30$, then the relative contribution from
secondary  non-Gaussianities are subdominant compared to the primordial non-Gaussianity. Alternatively, if WMAP data are used to
constrain that $f_{\rm NL} < 30$, then such a constraint must account for the covariances from secondary non-Gaussianities, especially those involving CMB lensing.

With Planck, non-Gaussianity estimates can
be extended to $l_{\rm max}\sim 2000$, but at such small angular scales, one finds a bias higher by a factor of more than 2 relative to
the lowest value of $f_{\rm NL}$ that can be reached with Planck (dot-dashed line). 
In return, if Planck data were to constrain  $f_{\rm NL}$ to be below $\sim$ 20, then such a constraint must account for
the confusion from secondary anisotropies to the ``optimal'' estimator of $f_{\rm NL}$,
since lensing non-Gaussianities produce a correction to $f_{\rm NL}$ with $f_{\rm bias} \sim 10$.

To account for secondary non-Gaussianities, one can modify 
existing ``optimal estimators'' for $f_{\rm NL}$ and jointly fit for
both the primordial non-Gaussianity and the secondary non-Gaussianities through a series of estimators
$\hat{S}_{\alpha}$ where $\alpha$ denotes the non-Gaussianity of interest with 
\begin{equation}
\hat{S}_{\alpha}=N_{\alpha,\beta}K_{\beta} \, ,
\end{equation} 
where
\begin{equation}
N_{\alpha,\beta} = \sum_{l_1 l_2 l_3} \frac{B^{\rm \alpha}_{l_1l_2l_3}B^{\rm \beta}_{l_1l_2l_3}}{\sigma^2(l_1,l_2,l_3)} \, .
\end{equation}
and $K_{\beta}$ refers to the set of non-Gaussianity parameters: 
$(f_{\rm NL}, b_{\rm ps}, A_{\rm SZ}, A_{\rm ISW})$. This method assumes
that one has a good model for $(l_1,l_2,l_3)$ dependence of secondary
bispectra $B^{\beta}_{l_1l_2l_3}$. Even if the point source covariance is small, the amplitude of the
point source confusion is generally unknown. Moreover, at $l < 1500$ many secondary bispectra such as the SZ effect
has a power-law behavior similar to the bispectrum of point sources. Thus, it would be necessary to determine the
amplitude $b_{\rm ps}$ from a joint fit.

Our suggestion that an accounting of secondary anisotropies is necessary for primordial non-Gaussian measurement is 
different from the general assumption in the literature that one can simply ignore the covariance between
primordial and secondary non-Gaussianities. This partly comes through, for example, 
the suggestion that primordial and point-source bispectra are orthogonal following
results from an exercise that involved jointly measuring non-Gaussian amplitudes $f_{\rm NL}$ and $b_{\rm ps}$ 
using a set of simulated maps in Ref.~\cite{Komatsu4} to study if there are biases in the estimators. 
However, this study used simulated non-Gaussian maps that did not include
any point sources with $b_{\rm ps}=0$. This sets the covariance  to be zero
and we believe this may have led to the wrong conclusion that there is no bias
in the optimal estimator for $f_{\rm NL}$ from unresolved point sources, though such a bias
is expected to be small, but non-negligible if $f_{\rm NL} \sim 1$.  Our conclusions are
consistent with some of the observations in Ref.~\cite{Smith}.

Here we have considered the confusion from secondary non-Gaussianities such as point sources and
those generated by CMB lensing. Additional contributions to the bispectrum exist with correlations between
SZ, ISW and point sources as they all trace the same large-scale structure at low redshifts.
Previous studies using the halo model to describe the  non-linear density field have shown correlations
such as between SZ and radio sources to be small \cite{Cooray2}, but since the bispectra in these cases are
of the form SZ-PS-PS$^{\rm nl}$, these bispectra may have a multipolar dependence in $(l_1,l_2,l_3)$
that is more aligned with the CMB primary bispectrum. In an upcoming paper \cite{Sarkar} we will discuss the
impact of such foreground bispectra due to correlations between CMB secondary anisotropies and
point sources. While our discussion has concentrated on a momentum independent non-Gaussianity parameter $f_{\rm NL}$, or the so-called
local type associated squeezed triangles, it is easy to generalize the calculation for more complex descriptions of $f_{\rm NL}(\bk_1,\bk_2,\bk_3)$ \cite{Babich}.
Due to differences in mode overlap, 
the exact momentum dependence will change the covariance contributions and the impact of secondary non-Gaussianities will be different 
between attempts to measure local $f_{\rm NL}$ and, for example, equilateral $f_{\rm NL}$.

Based on our calculations on the covariance between lensing and primary bispectra we have suggested 
a potential confusion for $f_{\rm NL}$ measurement in Planck data.
It is unlikely that our observation on the importance of secondary non-Gaussianities changes any of the current constraints 
on the non-Gaussianity parameter with WMAP data given that they mostly lead to $f_{\rm NL} \lesssim 100$ roughly.
The secondary non-Gaussianities, however, could impact 
the significance of any detections of primordial non-Gaussianity, especially if the
detection is marginally different from zero \cite{Yadav}.
For such studies, the exact significant of the detection should 
include an accounting of the secondary non-Gaussianity and the overlap with primordial
bispectrum in the ``optimal'' estimator used to establish $f_{\rm NL}$.

We thank Eiichiro Komatsu for a helpful communication. This work was supported by NSF CAREER AST-0645427. We acknowledge the use of CMBFAST by Uros Seljak and Matias Zaldarriaga \cite{Seljak}.

\end{document}